\documentclass{article}
\usepackage{frascatiphys}
\usepackage{graphicx}
\begin{document}
\title{ 
AXION-LIKE PARTICLES AND HIGH ENERGY ASTROPHYSICS
}
\author{
Giorgio Galanti\thanks{ \, gam.galanti@gmail.com}        \\
{\em INAF, Osservatorio Astronomico di Brera, Via E. Bianchi 46, I -- 23807 Merate, Italy}
}
\maketitle
\baselineskip=11.6pt
\begin{abstract}
Axion-like particles (ALPs) are light, neutral, pseudo-scalar bosons predicted by several extensions of the Standard Model of particle physics -- such as the String Theory -- and are supposed to interact primarily only with two photons. In the presence of an external magnetic field, photon-ALP oscillations occur and can produce sizable astrophysical effects in the very-high energy (VHE) band ($100 \, {\rm GeV} - 100 \, {\rm TeV}$). Photon-ALP oscillations increase the transparency of the Universe to VHE photons partially preventing the gamma-gamma absorption caused by the Extragalactic Background Light (EBL). Moreover, they have important implications for active galactic nuclei (AGN) by modifying their observed spectra both for flat spectrum radio quasars (FSRQs) and BL Lacs. Many attempts have been made in order to constrain the ALP parameter space by studying irregularities in spectra due to photon-ALP conversion in galaxy clusters and the consequences of ALP emission by stars. Upcoming new VHE photon detectors like CTA, HAWC, GAMMA-400, LHAASO, TAIGA-HiSCORE, HERD and ASTRI will settle the issue.
\end{abstract}
\baselineskip=14pt
\section{Introduction}
The detection of axion-like particles (ALPs) would represent a stunning development in  particle physics, since it would drive fundamental research towards a very specific direction in order to understand the laws responsible for the evolution of our Universe. Furthermore, implications of the detection of an ALP would be dramatic in astroparticle physics, since ALP interaction with photons would modify many aspects of the gamma-ray propagation: the transparency of photons propagating in magnetized media, the emission models of active galactic nuclei (AGN), and the stellar evolution. In the following, we concentrate on the consequences of ALPs for very-high energy (VHE) astrophysics with arising features which may be observed by the above new generation of the VHE gamma-ray observatories.

\section{Axion-like particles}
Many extensions of the Standard Model (SM) of elementary particle physics such as the String Theory predict the existence of axion-like particles (ALPs), which are very light, neutral, pseudo-scalar bosons\cite{alp2}. ALPs are a generalization of the {\it axion}, the pseudo-Goldstone boson related to the global Peccei-Quinn symmetry ${\rm U}(1)_{\rm PQ}$ proposed as a natural solution to the strong CP problem\cite{axionrev4}. While for the axion its interaction with fermions, two gluons, and two photons must be considered, and a strict relationship between the axion mass and the two-photon coupling constant exists, ALPs differ from the axion in this respect: i) ALPs are supposed to couple primarily only to  two photons while other interactions are discarded, ii) ALP mass $m_a$ and the two-photon coupling constant $g_{a \gamma \gamma}$ are unrelated parameters. Thus, the Lagrangian describing the ALP field $a$ reads 
\begin{equation}
\label{t1}
{\cal L}_{\rm ALP} = \frac{1}{2} \, \partial^{\mu} a \, \partial_{\mu} a - \, \frac{1}{2} \, m_a^2 \, a^2 + g_{a \gamma \gamma} \, a \, {\bf E} \cdot {\bf B}~,
\end{equation}
where ${\bf E}$ and ${\bf B}$ are the electric and magnetic components of the electromagnetic tensor $F^{\mu \nu}$. As far as the actual value of $m_a$ and $g_{a \gamma \gamma}$ is concerned, many bound have been derived as discussed in\cite{grExt} and also below in the text.

Considering a photon-ALP beam, we denote by $\bf B$ an external magnetic field and by $\bf E$ the propagating photon field in eq.\ref{t1}: because the mass matrix of the $\gamma - a$ system is off-diagonal, the propagation eigenstates differ from the interaction eigenstates. As a result, in the presence of an external $\bf B$ field, photon-ALP oscillations take place. Furthermore, from eq.\ref{t1} it follows that the only part of ${\bf B}$ which couples to $a$ is the transverse component ${\bf B}_T$ belonging to the plane containing $\bf E$ and orthogonal to the photon momentum ${\bf k}$. From eq.\ref{t1} it is possible to evaluate  the photon survival probability $P_{\gamma \to \gamma}$ representing the observable quantity: indeed $P_{\gamma \to \gamma}$ is related to the optical depth $\tau$ by the relation $P_{\gamma \to \gamma}=e^{-\tau}$\cite{dgr11}. In media magnetized by strong ${\bf B}$ such as inside the jet of AGN we must take into account also the one-loop QED vacuum polarization\cite{rs1988}.

\section{ALP impact in very-high-energy astrophysics}
The existence of ALPs would have many implications in VHE astrophysics in each environment where magnetic fields are sufficiently strong and/or the path inside a magnetized medium is long.
In this Section, we cursorily sketch the consequences of ALPs in these environments.

\subsection{Active galactic nuclei}
Active galactic nuclei (AGN) are powered by mass accreting onto supermassive black holes and in some cases they are characterized by the formation of two collimated relativistic oppositely oriented jets. An enormous amount of radiation is emitted from the inner regions. When one of the 
jets is occasionally pointing toward the Earth the AGN is called a blazar. Blazars are divided into two classes: flat spectrum radio quasars (FSRQs) and BL Lac objects (BL Lacs). FSRQs are more powerful and they are characterized by strong optical emission lines, while BL Lacs are less bright and they do not display significant emission lines.

\begin{figure}[htb]
    \begin{center}
        {\includegraphics[height=0.345\linewidth]{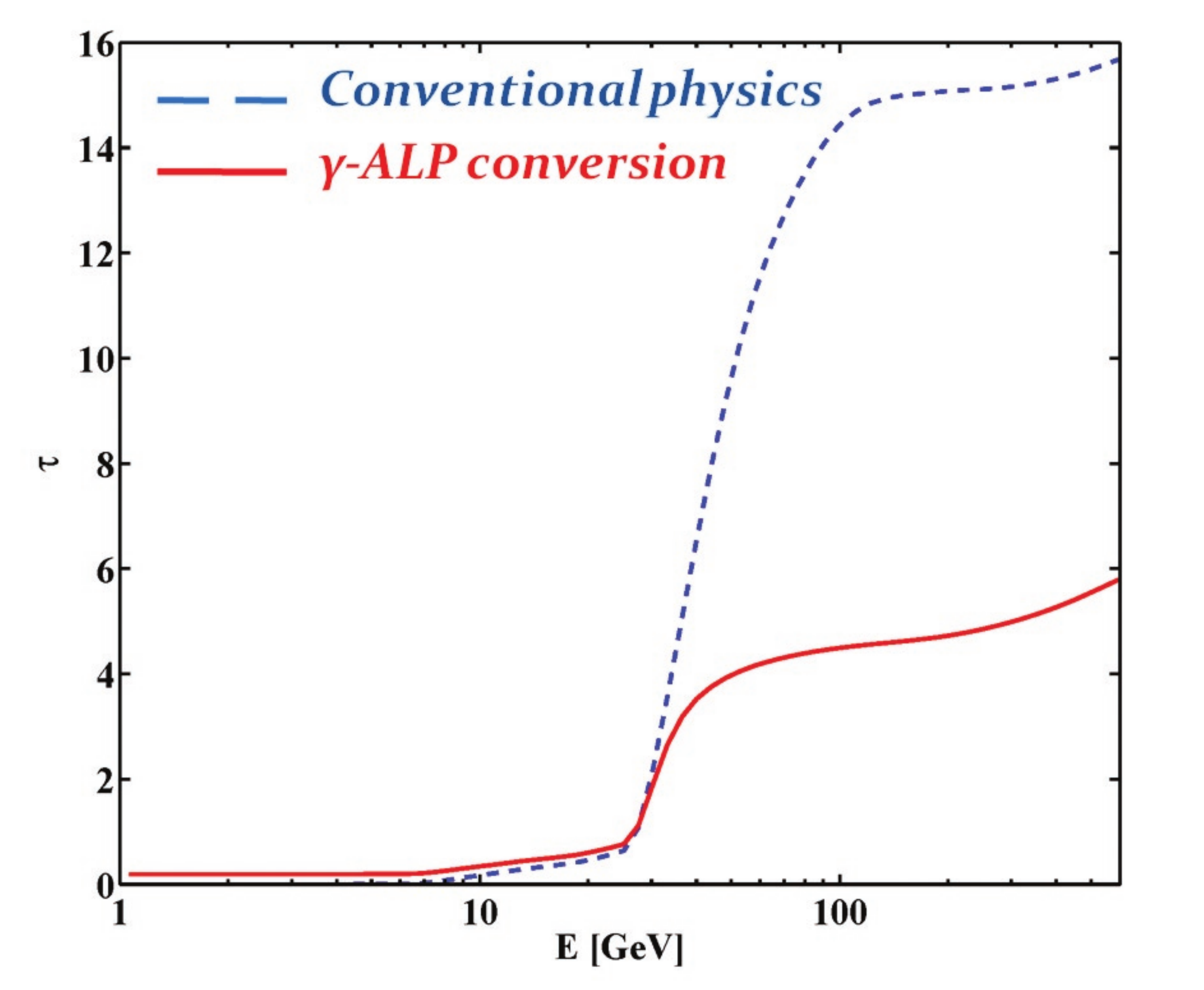}}
        {\includegraphics[height=0.35\linewidth]{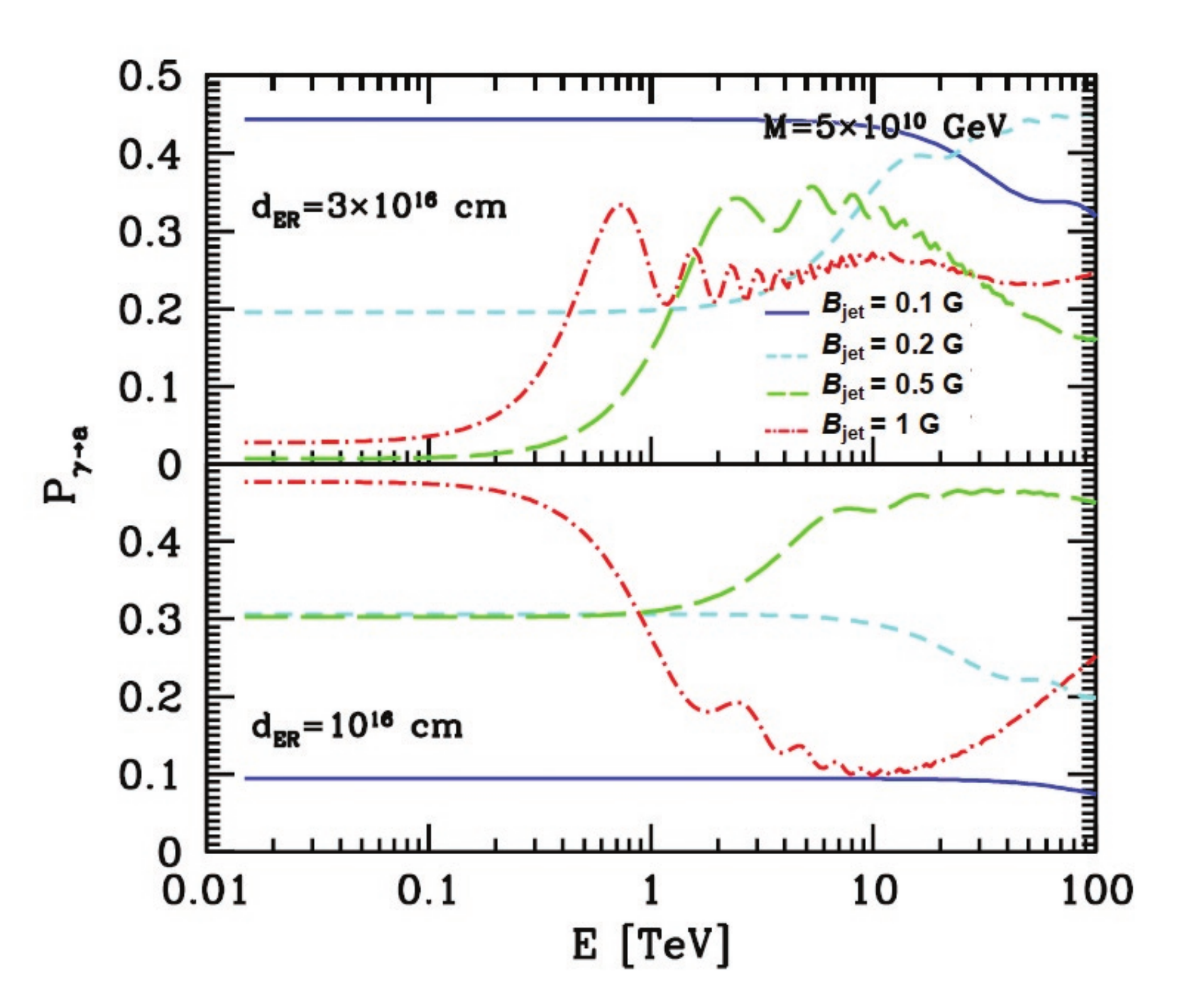}}
        \caption{\it Photon-ALP interaction in blazars. In the left panel, for FSRQs we report the BLR optical depth $\tau$ in the standard case and with photon-ALP interaction: $g_{a\gamma\gamma} \simeq 10^{-11} \, \rm GeV^{-1}$ and $m_a < {\cal O}(10^{-10} \, \rm eV)$\cite{trgb2012}. In the right panel, for BL Lacs we plot the photon-to-ALP conversion probability $P_{\gamma \to a}$ with $M \equiv 1/g_{a\gamma\gamma} = 5 \times 10^{10} \, \rm GeV$ and different choices of the emission distance and of the jet magnetic field $B_{\rm jet}$\cite{trg2015}.} 
\label{AGN}
    \end{center}
\end{figure}

In FSRQs the VHE photons produced at the jet base interact with the optical photons of the broad line region (BLR) thereby disappearing by producing an $e^+e^-$ pair: in the BLR the optical depth $\tau$ is so high that no photon with energies above $\sim 20 \, \rm GeV$ is  expected to leave the FSRQs. However, photons with energies up to $\sim 400 \, \rm GeV$ have been observed. In order to explain such a detection one is forced to place the emission region beyond the BLR in order to avoid absorption. If we want instead to place the emission region -- as in the standard AGN models -- not too far from the center, then  photon-ALP oscillations inside the jet magnetic field $B_{\rm jet} =  {\cal O}(1\, \rm G)$ can  be invoked in order to reduce the effective BLR optical depth\cite{trgb2012}. In the left panel of fig.\ref{AGN} we can observe the dramatic reduction of the optical depth $\tau$ in the presence of photon-ALP interactions as compared to the standard $\tau$: ALPs do not interact with BLR photons thereby increasing the effective VHE photon mean free path, thereby providing an explanation for photon emission well above $20 \, \rm GeV$ without invoking ad hoc blazar emission models. A complete and physically motivated spectral energy distribution (SED) is plotted in figs.8-10 of\cite{trgb2012}.

BL Lacs do not present absorption regions as FSRQs so that VHE photons can escape from the central region unimpeded. In any case, photon-ALP conversion inside the jet magnetic field $B_{\rm jet} = (0.1-1) \, \rm G$ allows for the production of ALPs already in the source. The fraction of photon/ALP produced strongly depends on the values of the emission distance from the centre, on $B_{\rm jet}$ and on $g_{a\gamma\gamma}$\cite{trg2015}. The right panel of fig.\ref{AGN} shows that the estimate for the produced ALPs is fairly large with a reasonable choice of the parameters.

\subsection{Galaxy clusters}
The emission of gamma-rays from the Perseus cluster -- possessing a magnetic field $B_{\rm clus}=1-10 \, \rm \mu G$ -- has been used in order to set constrains on $m_a$ and $g_{a \gamma \gamma}$ by studying the possible photon-to-ALP conversion inside $B_{\rm clus}$ and back conversion inside the magnetic field of the Milky Way. A generic photon-ALP conversion probability is plotted versus the energy $E$ in fig.7 of\cite{grSM} with arbitrary reference energy $E_{\rm ref}$. When the ALP mass effect at low energies is important (see the oscillatory region at low energy in fig.7 of\cite{grSM}) $P_{\gamma \to a}(E, m_a, g_{a\gamma\gamma}, B_{\rm clus})$ predicts spectral irregularities in observational data, which are related to the value of $m_a$ and $g_{a\gamma\gamma}$.
Since the observable spectrum $F_{\rm obs}$ is related to the intrinsic one $F_{\rm em}$ by $F_{\rm obs}= P_{\gamma \to \gamma}(m_a, g_{a\gamma\gamma}) F_{\rm obs}$, with $P_{\gamma \to \gamma}$ denoting the photon survival probability in the presence of photon-ALP interaction, the statistical {\it no preference} for photon-ALP conversion to fit data constrains ALP parameters to: $g_{a\gamma\gamma} < 5 \times 10^{-12} \, \rm GeV^{-1}$ for $5 \times 10^{-10} \, {\rm eV} < m_a < 5 \times 10^{-9} \, \rm eV$\cite{fermi2016}.

\subsection{Extragalactic space}
\begin{figure}[htb]
    \begin{center}
        {\includegraphics[height=0.34\linewidth]{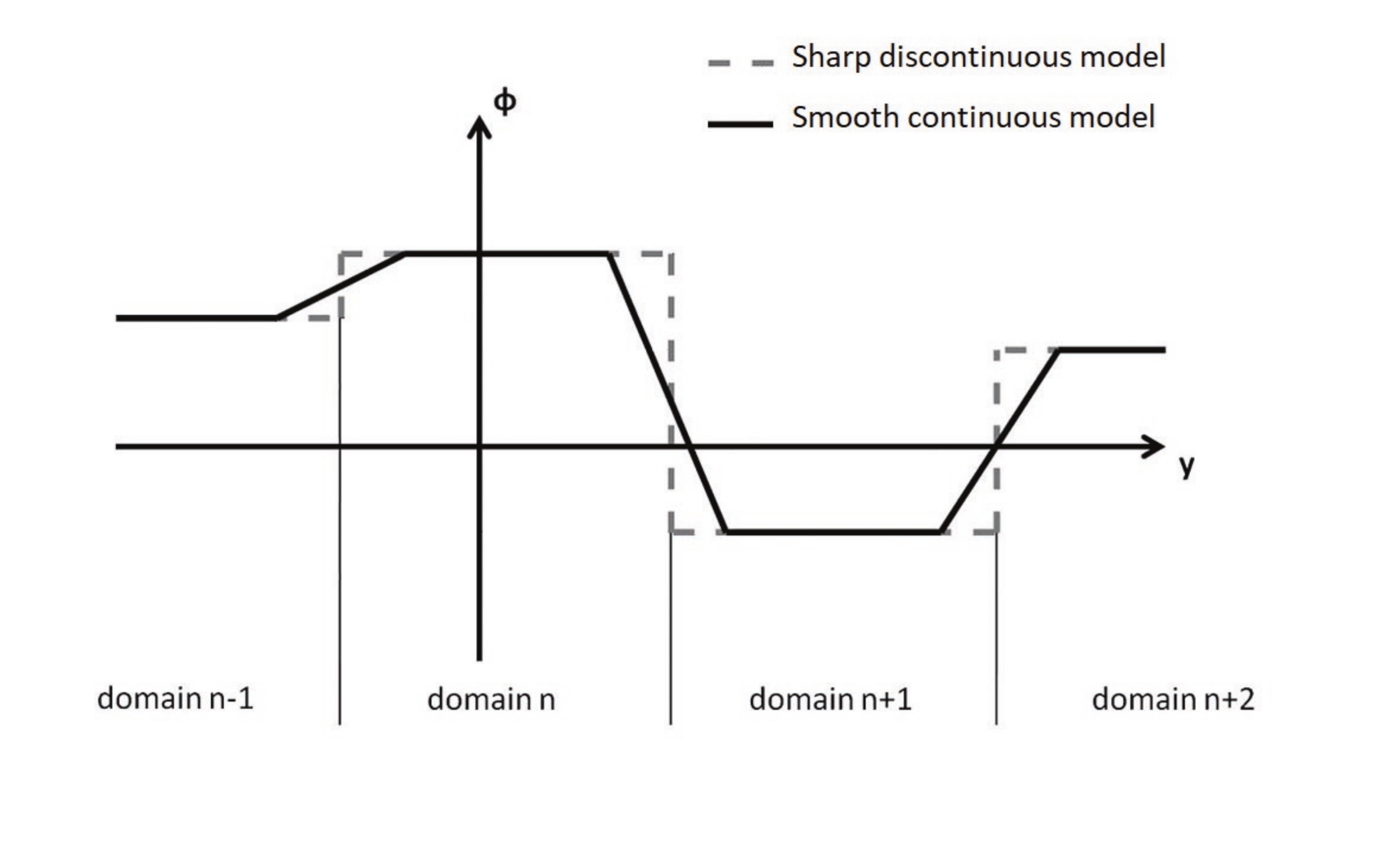}}
        {\includegraphics[height=0.35\linewidth]{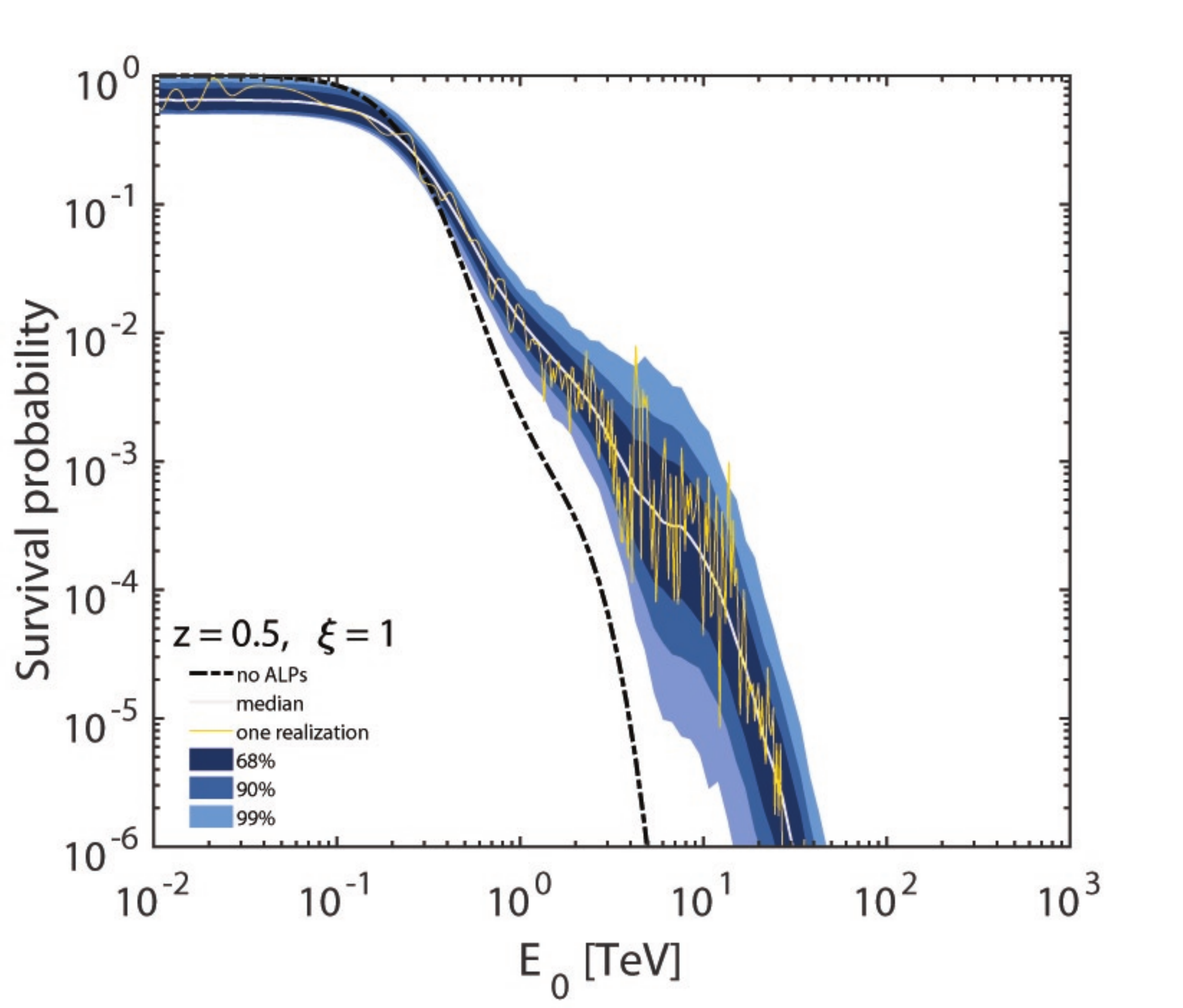}}
        \caption{\it Photon-ALP oscillations in extragalactic space. In the left panel, we plot the orientation angle $\phi$ of the transverse component of ${\bf B}_{\rm ext}$ crossing from a domain to the next in the domain-like smooth-edges (DLSME) model\cite{grSM}. In the right panel, we report the photon survival probability $P_{\gamma \to \gamma}$ versus the observed energy $E_0$ in the standard case and as modified by photon-ALP interaction for a redshift $z=0.5$ and $\xi=1$: a single realization of $P_{\gamma \to \gamma}$ and its statistical properties are plotted. See\cite{grExt} for details.} 
\label{extr}
    \end{center}
\end{figure}

The propagation of the photon-ALP beam in extragalactic space is affected by the morphology and strength of the extragalactic magnetic field ${\bf B}_{\rm ext}$ which has large uncertainties in the range $10^{-7} \, {\rm nG} < B_{\rm ext} < 1.7 \, \rm nG$ on the scale ${\cal O}(1 \, \rm Mpc)$. However, models contemplating galactic outflows especially from  dwarf galaxies are generally believed to be the source of ${\bf B}_{\rm ext}$. Accordingly, ${\bf B}_{\rm ext}$ turns out to have a domain-like structure with $B_{\rm ext}={\cal O}(1 \, \rm nG)$ and a coherence scale ${\cal O}(1 \, \rm Mpc)$. In extragalactic space photons are absorbed by scattering off the photons of the extragalactic background light (EBL) -- which is the infrared/optical/ultraviolet radiation emitted by galaxies during the cosmic evolution -- thereby producing an $e^+e^-$ pair\cite{dgr13,gtpr}. This process is the analogous of the photons in FSRQs interacting with those of the BLR: thus, photon-ALP oscillations can increase the VHE photon mean free path enhancing the Universe transparency\cite{dgr11}. In addition, photon dispersion on the cosmic microwave background (CMB) plays a crucial role above $15 \, \rm TeV$, where the sharp discontinuous domain-like model for ${\bf B}_{\rm ext}$ gives unphysical results about the calculation of the photon survival probability $P_{\gamma \to \gamma}$ and -- as reported in the left panel of fig.\ref{extr} -- a model which smoothly interpolates ${\bf B}_{\rm ext}$ from one domain to the next (with the orientation angle $\phi$ of the transverse component of ${\bf B}_{\rm ext}$ becoming a continuous function) is compelling\cite{grSM}.
As reported in the right panel of fig.\ref{extr}, we observe that, even in the presence of CMB photon dispersion, $P_{\gamma \to \gamma}$ with photon-ALP oscillations at work is drastically increased as respect to the conventional physics prediction\cite{grExt}. In the right panel of fig.\ref{extr}, we plot $P_{\gamma \to \gamma}$ for an hypothetical source at redshift $z=0.5$ and $\xi \equiv (B_{\rm ext}/{\rm nG})(g_{a\gamma\gamma}10^{11} \, \rm GeV)=1$. The behavior of $P_{\gamma \to \gamma}$ in the presence of photon-ALP oscillations depends on the choice of $\xi$ and $P_{\gamma \to \gamma}$ gets increased more and more as the redshift grows\cite{grExt}.

\subsection{Milky Way and total effect}
\begin{figure}[htb]
    \begin{center}
        {\includegraphics[height=0.35\linewidth]{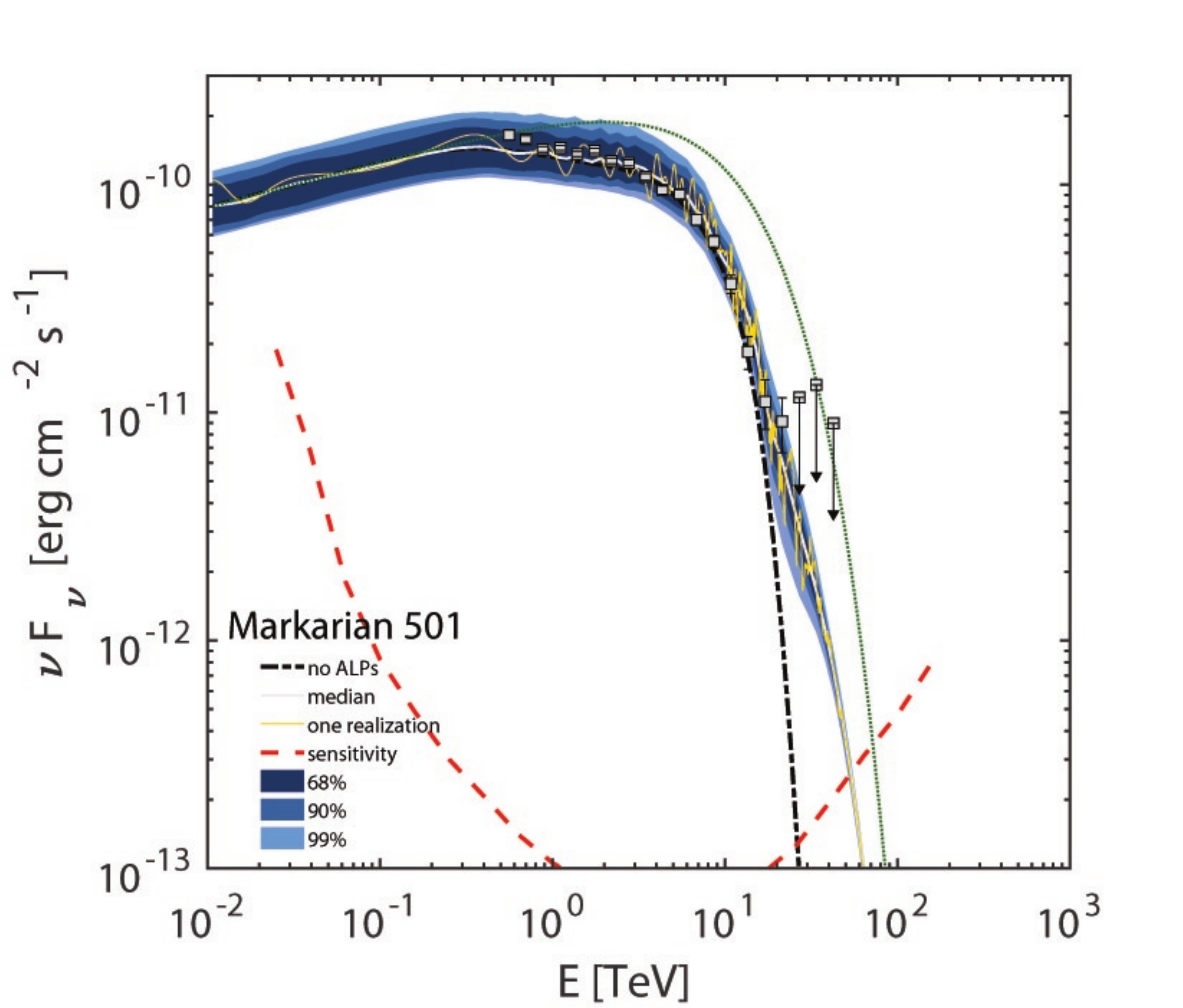}}
        {\includegraphics[height=0.35\linewidth]{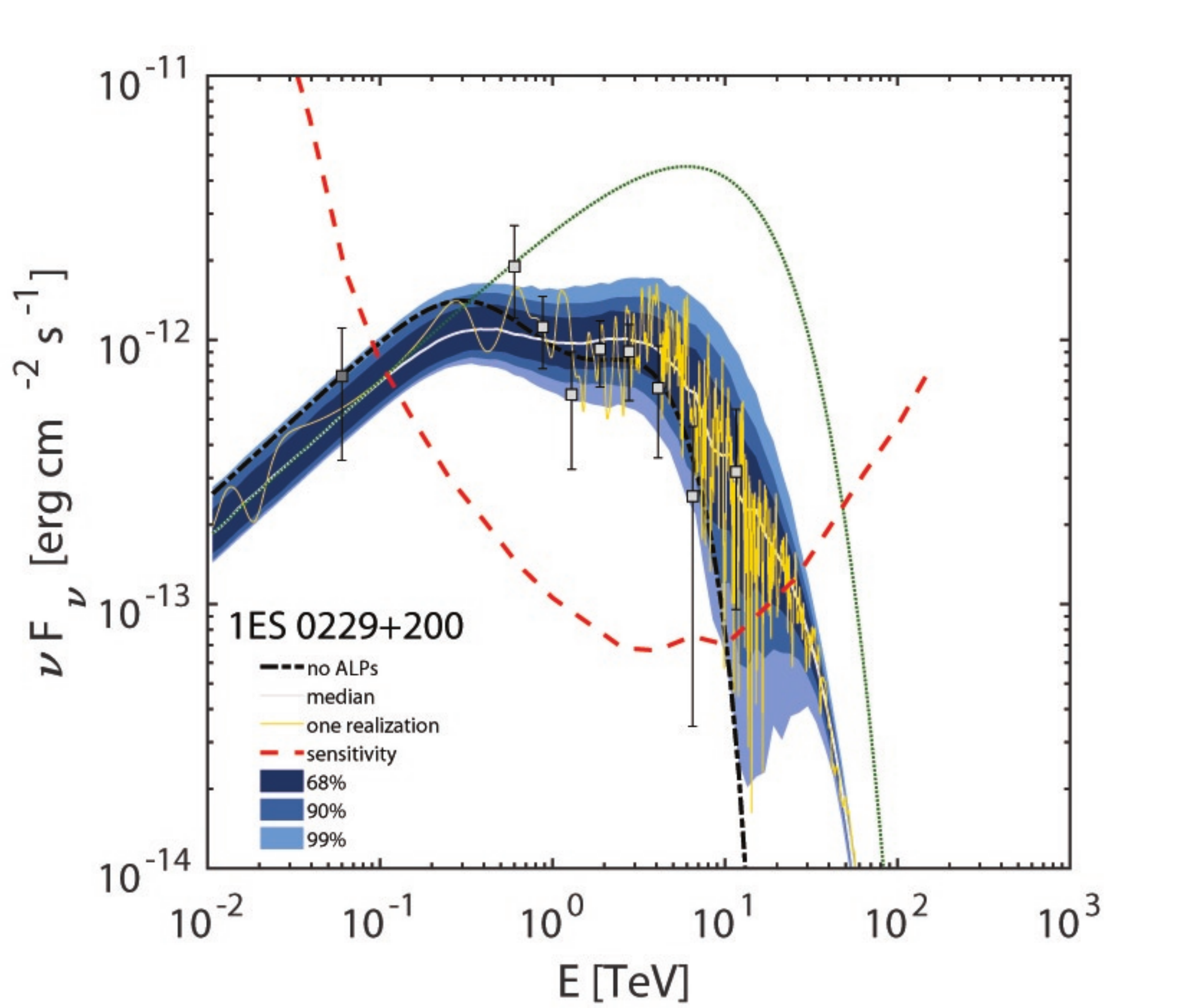}}
        \caption{\it Behavior of the observed SED of Markarian 501 (left panel) and 1ES 0229+200 (right panel) versus the observed energy $E$ in the standard case and in the 
        ALP scenario for reasonable values of $m_a$ and of $g_{a\gamma\gamma}$\cite{gtre} as discussed in previous sections.} 
\label{tot}
    \end{center}
\end{figure}

Only the regular component of the Milky Way magnetic field $B_{\rm MW} \simeq 5 \, \mu{\rm G}$ with coherence length $l_{\rm coh} \simeq 10 \, \rm kpc$ gives a sizable contribution to the photon-ALP conversion: in any case, detailed sky maps of $B_{\rm MW}$ exist. By combining the photon/ALP propagation in the several magnetized environments crossed by the photon-ALP beam (the jet, the extragalactic space, the Milky Way) it is possible to obtain the observed SED for BL Lacs. In fig.\ref{tot} we plot the SED of Markarian 501 (left panel) and 1ES 0229+200 (right panel) for reasonable values of $m_a$ and of $g_{a\gamma\gamma}$ as discussed above.
From fig.\ref{tot} we see that conventional physics hardly explains the highest energy points in the spectra of Markarian 501 and of 1ES 0229+200, while the model including photon-ALP oscillations naturally matches the data. The photon-ALP interaction predicts features in BL Lac spectra (an energy-dependent oscillatory behavior and a photon excess above $20 \, \rm TeV$) that may be tested by the above new generation of the VHE gamma-ray observatories\cite{gtre}.

\subsection{Main sequence and evolved stars}
ALPs can be produced in the Sun via Primakoff scattering $p + \gamma \to p + a$, where $p$ denotes a proton or a charged particle. The CAST experiment uses the fact that ALPs can then be reconverted back to photons inside the magnetic field of a decommissioned magnet of the LHC. However, no signal has been detected -- which gives a firm bound about $m_a$ and $g_{a\gamma\gamma}$: $g_{a\gamma\gamma} < 0.66 \times 10^{-10} \, \rm GeV^{-1}$ for $m_a < 0.02 \, \rm eV$\cite{CAST}.

ALPs similarly produced in main-sequence stars are a source of stellar cooling, which modifies stellar evolution as a function of $m_a$ and $g_{a\gamma\gamma}$. As a consequence, also globular cluster star content gets accordingly modified: from a comparison with observational data bounds on ALP parameters can be derived as $g_{a\gamma\gamma} < 0.66 \times 10^{-10} \, \rm GeV^{-1}$\cite{globclu}.

\section{Discussion and conclusions}
We have described some consequences of photon-ALP interaction in astrophysics but the list would be much longer: search for spectral irregularities of point sources in galaxy clusters in the X-ray energy band\cite{spectrX}; spectral distortions of the continuum thermal emission ($T \sim 2-8 \, \rm keV$) of galaxy clusters\cite{thermal}; study of the unexpected spectral line at $3.55 \, \rm keV$ as dark matter decay into ALPs and subsequent conversion to photons\cite{DMdecay}; in the VHE band hints about a better description of AGN spectral indices by introducing ALPs\cite{troitsky14,grdb}; search for a diffuse flux of photons coming from ALP-to-photon back-conversation concomitant with neutrino production in the extragalactic space\cite{laha}.

In conclusion, we stress that many of the previous effects arise with the same choice of the ALP model parameters ($m_a$, $g_{a\gamma\gamma}$): this fact may be a possible first hint of the existence of an ALP. In any case, other possibilities that partially mimic ALP effects exist and have been compared in\cite{gtl}. Astrophysical new data from the new generation of $\gamma$-ray observatories CTA, HAWC, GAMMA-400, LHAASO, TAIGA-HiSCORE, HERD and ASTRI will provide a check of the scenarios outlined above.

\section*{Acknowledgements}
I thank M. Roncadelli and F. Tavecchio for a careful reading of the manuscript. I acknowledge contribution from the grant INAF CTA--SKA, ``Probing particle acceleration and $\gamma$-ray propagation with CTA and its precursors" and INAF Main Stream project ``High-energy extragalactic astrophysics: toward the CTA''.


\begin{thebibliography}{99}
\bibitem{alp2} A. Ringwald, Phys. Dark Univ. {\bf 1}, 116 (2012).

\bibitem{axionrev4} D. J. E.   Marsch, Phys. Rep. {\bf 643}, 1 (2016).

\bibitem{grExt} G. Galanti and M. Roncadelli, J. High Energy Astrophys. {\bf 20}, 1 (2018).

\bibitem{dgr11} A. De Angelis, G. Galanti and M. Roncadelli, Phys. Rev. D {\bf 84}, 105030 (2011).

\bibitem{rs1988} G. G. Raffelt and L. Stodolsky, Phys. Rev. D {\bf 37}, 1237 (1988).

\bibitem{trgb2012}  F. Tavecchio, M. Roncadelli, G. Galanti and G. Bonnoli, Phys. Rev. D {\bf 86}, 085036 (2012).

\bibitem{trg2015} F. Tavecchio, M. Roncadelli and G. Galanti, Physics Letters B {\bf 744}, 375 (2015).

\bibitem{grSM} G. Galanti and M. Roncadelli, Phys. Rev. D {\bf 98}, 043018 (2018).

\bibitem{fermi2016} M. Ajello {\it et al}., Phys. Rev. Lett. {\bf 116}, 161101 (2016).

\bibitem{dgr13} A. De Angelis, G. Galanti and M. Roncadelli, Mon. Not. R. Astron. Soc. {\bf 432}, 3245 (2013).

\bibitem{gtpr} G. Galanti, F. Tavecchio, F. Piccinini and M. Roncadelli, arXiv:1905.13713 (2019).

\bibitem{gtre} G. Galanti, F. Tavecchio, M. Roncadelli and C. Evoli, Mon. Not. R. Astron. Soc. {\bf 487}, 123 (2019).

\bibitem{CAST} V. Anastassopoulos {\it et al.}, Nat. Phys. {\bf 13}, 584 (2017).

\bibitem{globclu} A. Ayala {\it et al.}, Phys. Rev. Lett., {\bf 113}, 1302 (2014).

\bibitem{spectrX} J. P. Conlon, F. Day, N. Jennings, S. Krippendorf and M. Rummel, JCAP {\bf 1707}, 005 (2017).

\bibitem{thermal} J. P. Conlon, M. C. D. Marsh and A. J. Powell, Phys. Rev. D {\bf 93}, 123526 (2016).

\bibitem{DMdecay} J. Jaeckel, J. Redondo and A. Ringwald, Phys. Rev. D {\bf 89}, 103511 (2014).

\bibitem{troitsky14} G. I. Rubtsov and S. V. Troitsky, JETP Lett., {\bf 100}, 355 (2014).

\bibitem{grdb} G. Galanti, M. Roncadelli, A. De Angelis and G. F. Bignami, arXiv:1503.04436 (2015).

\bibitem{laha} H. Vogel, R. Laha and M. Meyer, arXiv:1712.01839 (2017).

\bibitem{gtl} G. Galanti, F. Tavecchio, M. Landoni, arXiv:1911.09056 (2019).

\end{thebibliography}
\end{document}